\documentstyle[12pt,epsf,sprocl]{article}
\def\lsi{\raise0.3ex\hbox{$<$\kern-0.75em\raise-1.1ex\hbox{$\sim$}}}
\def\gsi{\raise0.3ex\hbox{$>$\kern-0.75em\raise-1.1ex\hbox{$\sim$}}}
\newcommand{\lsim}{\mathop{\lsi}}

\newcommand{\ba}{\begin{eqnarray}}
\newcommand{\ea}{\end{eqnarray}}

\newcommand{\be}{\begin{equation}}
\newcommand{\ee}{\end{equation}}

\begin{document}
\begin{flushright}
\end{flushright}
\title{Baryogenesis at the Electroweak Phase Transition for a
 SUSY Model\\ with a Gauge Singlet\footnote{To appear in the 
Proceedings of SEWM2000, Marseille, June 14-17, 2000 }}

\author{S.~J.~Huber}
\address{Bartol Research Institute, University of Delaware, Newark, DE 19716\\
E-mail: shuber@bartol.udel.edu}
\author{M.~G.~Schmidt}
\address{Institut f\"ur Theoretische Physik, Universit\"at Heidelberg,
69120 Heidelberg, Germany\\
E-mail: m.g.schmidt@thphys.uni-heidelberg.de}

\maketitle

\abstract{
SUSY models with a gauge singlet easily allow for a strongly 
first order electroweak phase transition (EWPT). We discuss the
wall profile, in particular transitional CP violation during the EWPT.
We calculate CP violating source terms for the charginos in
the WKB approximation and solve the relevant transport
equations to obtain the generated baryon asymmetry.
}
\section{Introduction}
The ingredients of electroweak baryogenesis, a first order phase
transition, CP violation and baryon number violation can
be used to work out theoretically a large asymmetry in a very concrete
way. They can be tested in experiments at the electroweak scale and in
lattice simulations. The standard model does not provide a phase
transition with the present bounds on the Higgs mass and it also does
not contain strong enough CP violation. This is different in supersymmetric
variants of electroweak models. In the MSSM there is a (rather small )
corner left - with the lightest Higgs mass above 100 GeV and $stop_R$
mass slightly below $m_{top}$ - there one can produce sizable baryon
asymmetry\cite{litestop}. In NMSSM type supersymmetric models with an additional
singlet there is much more parameter space for successful
baryogenesis\cite{PDFM,HS2,HS3}.
\section{The model}
\begin{figure}[t] 
\begin{picture}(100,90)
\put(-70,-332){\epsfxsize14cm \epsffile{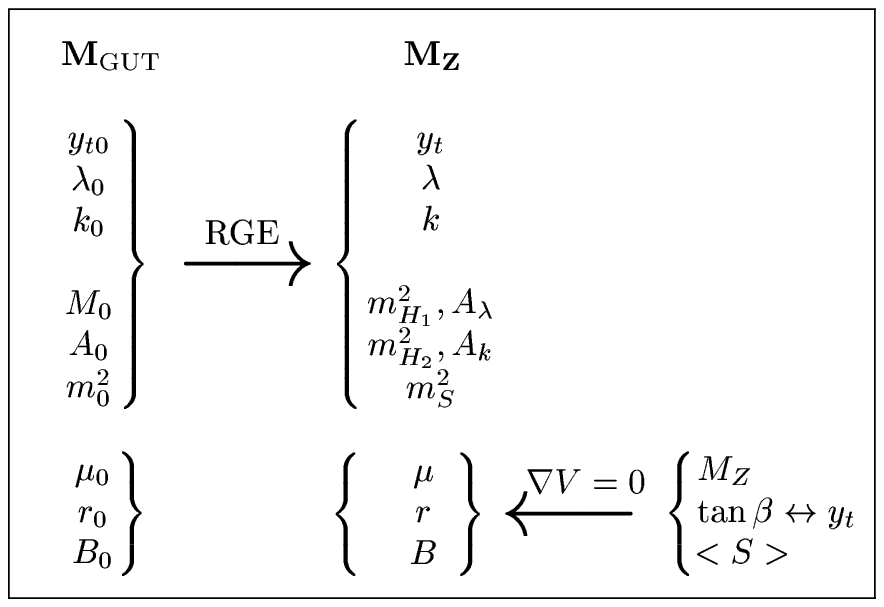}}
\put(120,-262){\epsfxsize12cm \epsffile{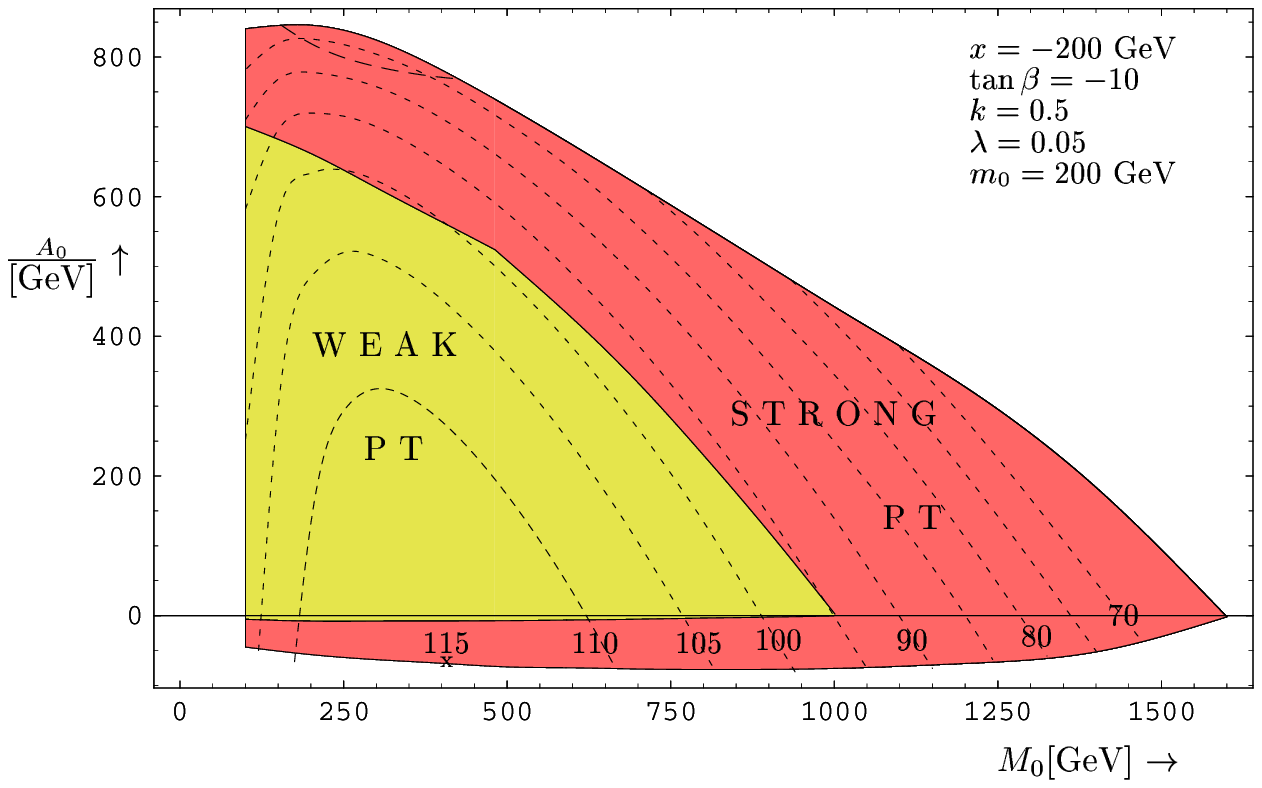}}
\put(100,130){(a)} 
\put(310,130){(b)} 
\end{picture} 
\caption{
(a): Sketch of our procedure to determine the weak
scale parameters from the GUT parameters.
(b): Scan of the $M_0$-$A_0$ plane for a set of $(x,\tan\beta,k)$: In the red (yellow) areas the
PT is strongly (weakly) first order, i.e.~$v_c/T_c>1$ ($v_c/T_c<1$). Dotted lines
are curves of constant mass of the lightest CP-even Higgs boson. In the region
above the dashed line the lightest Higgs is predominantly a singlet.
}
\label{f_1}
\end{figure}

How one should go ``beyond'' is a completely open question. Supersymmetric
models are very promising but (still) not checked by experiments. We
discuss a SUSY model which contains besides the fields of the MSSM a
gauge singlet with the superpotential (``{ NMSSM}'')\cite{NMSSM}
\begin{equation}
W=\mu H_1H_2 +\lambda SH_1H_2+\frac{k}{3}S^3+rS 
\end{equation}
and (universal) soft SUSY breaking terms 
\begin{equation} 
{\cal L}_{\rm soft}\!=\!\lambda A_{\lambda}SH_1H_2+\frac{k}{3}A_kS^3+
Y_eA_e\tilde e^c\tilde lH_1+Y_dA_d\tilde d^c\tilde qH_1
+Y_uA_u\tilde u^c\tilde qH_2+\!\mbox{\small h.c.}
\end{equation}
Our final parameters are $$ \tan\!\beta, x, \lambda, k, M_0, A_0,
m_0^2, $$ where $\tan\!\beta$, {${x=<S>}$}, $\lambda$ and $k$ are fixed at
the electroweak scale, and $M_0$, $A_0$ and $m_0^2$ at the GUT scale.
The renormalisation group procedure is indicated in fig.~\ref{f_1}(a).

Already at the tree level there are terms in the potential of $\phi^3$-type\cite{PDFM,HS2},
\begin{equation}
(\lambda\mu^*S+\mbox{h.c.})(|H_1^0|^2+|H_2^0|^2)+
(\lambda A_{\lambda}SH_1^0H_2^0
+\frac{k}{3}A_kS^3+\mbox{h.c.}).
\end{equation}

Adding the usual 1-loop temperature dependent terms we can discuss
minima of the thermal potential and given the more general form of the
NMSSM (1) we can find a bright range of parameters where $<S>\sim
<H_{1,2}>$, and where the effective $\phi^3$-term is large enough to
produce a strongly first order phase transition. In fig.~\ref{f_1}(b) we show
an example of a scan in the $M_0$-$A_0$ plane, where a strongly
first order phase transition happens for Higgs masses up to 115 GeV\cite{HS2,HS3}. 

\section{CP-violating bubble walls}
We solve the equations of motion for the Higgs and singlet fields
\begin{eqnarray}
 &&H^0_{1,2}=\bar{h}_{1,2}e^{i\theta_{1,2}},\quad \theta= \theta_1+\theta_2, 
             \quad h=\sqrt{|H_1|^2+|H_2|^2}\\
&&S=n+ic,  \quad s=|S|,
\end{eqnarray}
to obtain the profile of the bubble wall \cite{HS3}. (See also the contribution 
of P.~John to these proceedings \cite{P}). CP violation leading
to non-vanishing $\theta$ and $c$ can be induced explicitly in the
parameters of the Higgs potential, or spontaneously. 

\begin{figure}[t] 
\begin{picture}(100,90)
\put(-65,-330){\epsfxsize13cm \epsffile{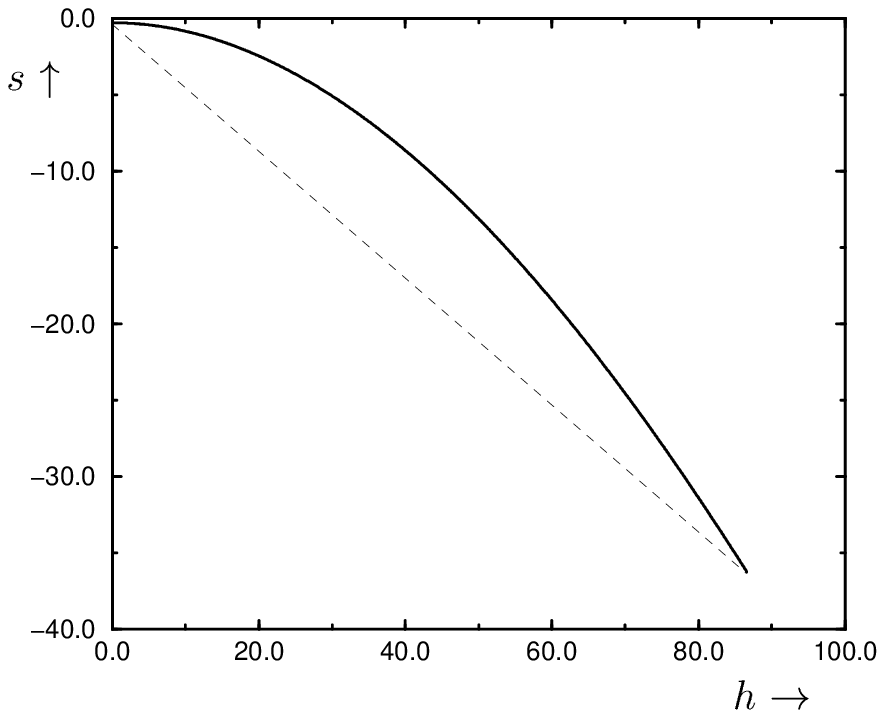}}
\put(188,-79){\epsfxsize7.7cm  \epsffile{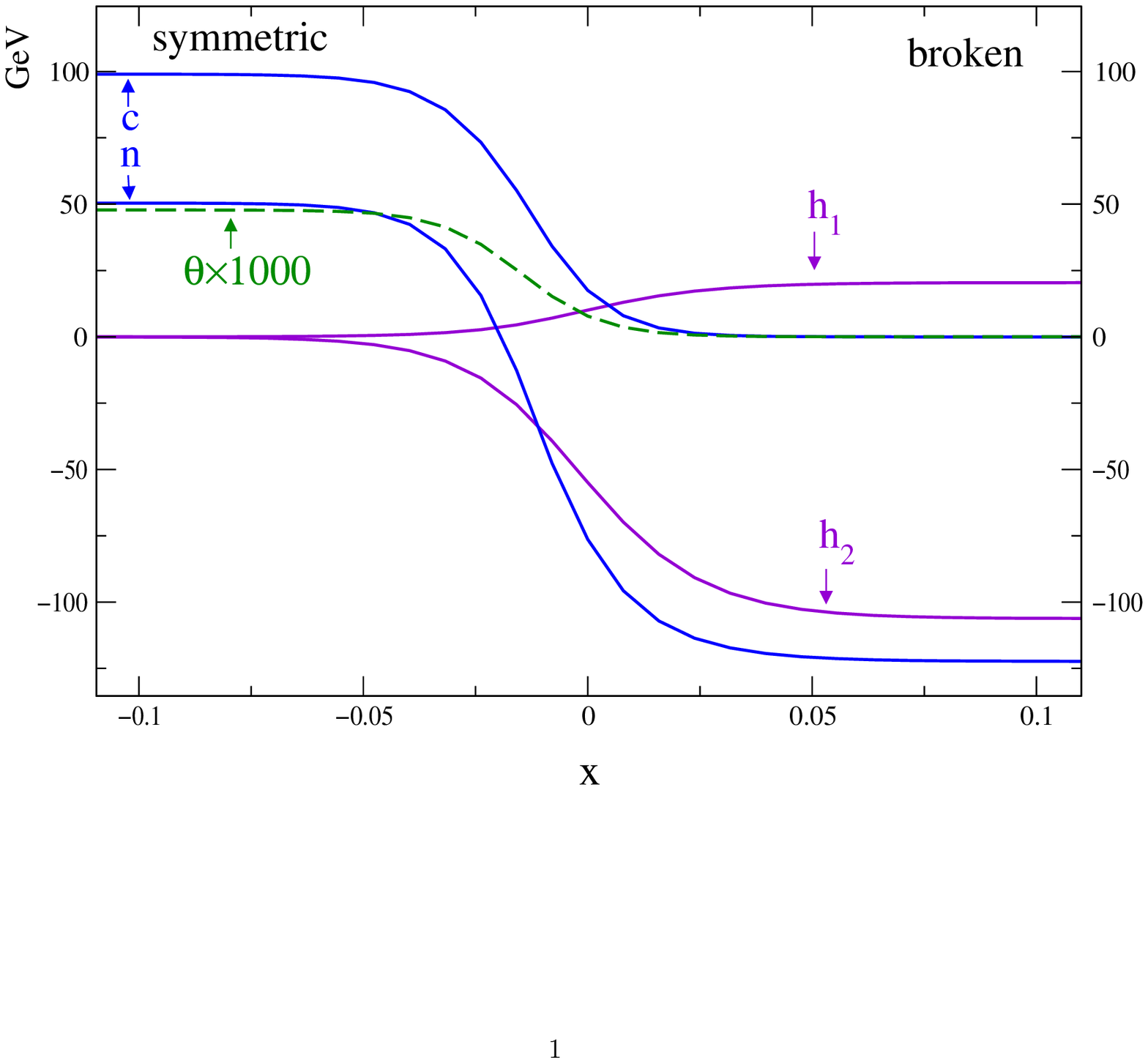}}
\put(120,120){(a)} 
\put(300,120){(b)} 
\end{picture} 
\caption{
(a): Example of a trajectory in the $h$-$s$ plane (solid line) and the straight
connection between the symmetric and the broken minimum (dashed line).
(b): Transitionally CP-violating bubble wall profile for some parameter set.
$x$ is the position variable. (All units in GeV.) 
}
\label{f_2}
\end{figure}

In the NMSSM 
there is the possibility of CP violation which is only present during the
phase transition (transitional CP violation) \cite{John4}. It provides large CP violation
for baryon number production, without generating large electric dipole
moments for the electron and neutron.
Fig.~\ref{f_2} shows two examples of bubble wall shapes in the NMSSM
for parameter sets given in ref.~\cite{HS3}.
\section{WKB approximation and dispersion relations}
For thick bubble walls, $L_w\gg 1/T$ the dispersion relations
of particles moving in the background of the Higgs profile can 
be reliably calculated in the WKB approximation \cite{CJK1}.
In the NMSSM we find $3/T<L_w<20$ \cite{HS3}. 
To order $\hbar$ the dispersion relations of charginos and stops
contain CP violating terms.  These enter as source terms in the
Boltzmann equations for the (particle-antiparticle) chemical 
potentials and fuel the creation of a baryon asymmetry through
the weak sphalerons in the hot phase.

In the NMSSM (like in the MSSM) the dominant source for
baryogenesis comes from the charginos with the mass matrix
\begin{equation}
{\cal L}= \cdots+
(i\tilde W^-,\tilde h_1^-)\left(\begin{array}{cc} 
M_2&g_2(H_2^0)^* \\ g_2(H_1^0)^* &\mu+\lambda S 
\end{array}\right) {i\tilde W^+ \choose \tilde h_2^+}
\end{equation}
Diagonalizing this mass matrix via 
$\bf M=VM_DU^{\dagger}$,
and solving the chargino Dirac equation in the WKB approximation,
we find 
two CP violating contributions for the dispersion relations
of chargino particles and antiparticles \cite{HS3}
$$ E =(\vec{p}^2+m^2)^{1/2}\pm\Delta  E$$
\begin{equation} \label{dis}
\Delta E={\rm sgn}(p_z)\frac{\theta_2'+\delta'\sin^2(b)-\gamma'\sin^2(a)}
{2\sqrt{\vec{p}^2+m_2^2}}\left(\sqrt{p_z^2+m_2^2}-|p_z|\right)
-\frac{p_z \gamma'\sin^2(a)}{\sqrt{\vec{p}^2+m_2^2}}
\end{equation}
with 
\begin{eqnarray}
\tan (2a)&=&\frac{2g_2|(M_2H_2^0)^*+H_1^0(\mu+\lambda S)|}
{|M_2|^2+g_2^2|H_1^0|^2-g_2^2|H_2^0|^2-|\mu+\lambda S|^2},\\
\gamma&=&\mbox{arg}\left((M_2H_2^0)^*+H_1^0(\mu+\lambda S)\right),
\end{eqnarray}
and $\tan(2\beta)$ and $\delta$ obtained by exchange of $H_1$ 
and $H_2$ being the parameter of the
diagonalization. As pointed out very recently in ref.\cite{CJK2}  one should better
use the kinetic momentum $p_{\rm kin}=m~\partial E/\partial p$ instead of the canonical
momentum in the semiclassical limit leading to the Boltzmann equations. We then
obtain a dispersion relation like (\ref{dis}) but with the last term omitted
and the factor accompanying the $(\theta'...)$ bracket changed to 
$m^2/2(p_{\rm kin}^2+m^2)$. $\Delta E$ is now totally symmetric under the 
exchange of $H_{1,2}$. This would destroy the most prominent 
term proportional to $(H_1'H_2 -H_2'H_1)$ in the older work on the MSSM.

\section{Diffusion  Equations and Application to the NMSSM}
We treat the Boltzmann equations for the transport of quasi-classical 
particles with dispersion relations discussed above,
\begin{equation}
d_tf_i=(\partial_t+\dot{\vec{x}}\cdot\partial_{\vec{x}}+
\dot{\vec{p}}\cdot\partial_{\vec{p}})f_i={\cal C}_i[f],
\end{equation}
in the fluid approximation \cite{CJK1}
\begin{equation} 
f_i(\vec{x},\vec{p},t)=\frac{1}{e^{\beta(E_i-v_ip_z-\mu_i)}\pm 1}.
\end{equation}
looking for a stationary solution, where $\bar z=z-v_wt$,
expanding in the perturbations and in $v_w$,  
averaging over $p_z$ with $p_z$, 1, 
taking the difference of particle and anti-particle chemical potentials,
one finds \cite{CJK1}
\begin{eqnarray}
-\kappa_i(D_i\mu_i''+v_w\mu_i')+\sum_p\Gamma_p^d\sum_j\mu_j=S_i,
\nonumber\\
S_i=\frac{D_iv_w}{\langle p_z^2/E_0 \rangle_0}
\langle p_z\Delta E_i'\rangle'-\sum_p\Gamma_p^d
\langle\Delta E_{{\rm sp},p}\rangle, 
\end{eqnarray}
with diffusion constants $D_i=\kappa_i\langle p_z^2/E_0\rangle_0^2/
(\bar p_z^2\Gamma_i^e)$, statistical factors $k_i$,
interaction rates $\Gamma$, and  CP-violating source terms $S_i$.

In the NMSSM, the relevant interactions are
\begin{eqnarray}
{\cal L_{\rm int}}&=&y_t t^cq_3H_2+y_t \tilde t^cq_3\tilde h_2 +
y_t t^c\tilde q_3\tilde h_2 - y_t\mu \tilde t^{c*}\tilde q_3^*H_1
+y_tA_t\tilde t^c\tilde q_3H_2
\nonumber\\
&& +\lambda \tilde s \tilde h_1H_2+
\lambda \tilde s \tilde h_2 H_1 +\mbox{h.c.}
\end{eqnarray}
and the supergauge interactions (in equilibrium!), the higgsino helicity
flips (from $\mu\tilde h_1 \tilde h_2$), the Higgs and axial top number
violation in the broken phase, and the strong sphalerons. The resulting
interaction terms in the diffusion equations are \cite{HS3}
\begin{eqnarray}
(\Gamma_y+\Gamma_{yA})(\mu_{H_2}+\mu_{Q_3}+\mu_T),\ \Gamma_{y\mu}(\mu_{H_1}-\mu_{Q_3}-\mu_T), 
\ \Gamma_{\lambda}(\mu_{\tilde s}+\mu_{H_1}+\mu_{H_2}),\nonumber \\
\Gamma_{ss}(2\mu_{Q_3}+2\mu_{Q_2}+2\mu_{Q_1}+\mu_T+\mu_B+\mu_C+\mu_S+\mu_U+\mu_D),
\nonumber\\
\Gamma_{hf}(\mu_{H_1}+\mu_{H_2}), \quad
\Gamma_m(\mu_{Q_3}+\mu_T), \quad \Gamma_{H_1}\mu_{H_1}, \quad
\Gamma_{H_2}\mu_{H_2}.
\end{eqnarray}

We obtain a reduced set of diffusion equations for the chemical potentials
$\mu_{Q_3}$, $\mu_T$, $\mu_{H_1}$, $\mu_{H_2}$ and $\mu_{\tilde s}$;
e.g.~for $\mu_{H_1}$, $\mu_{H_2}$ and $\mu_{\tilde s}$ they read \cite{HS3}
\begin{eqnarray}
-k_{H_1}{\cal D}_{H_1}\mu_{H_1}+
6\Gamma_{y\mu}[\mu_{H_1}-\mu_{Q_3}-\mu_T]+
2\Gamma_{\lambda}[\mu_{\tilde s}+\mu_{H_1}+\mu_{H_2}]
\hspace{1.4cm}\nonumber\\
+2\Gamma_{hf}(\mu_{H_1}+\mu_{H_2})+2\Gamma_{H_1}\mu_{H_1}=S_{H_1}
\hspace{.4cm}\\[.1cm]  
-k_{H_2}{\cal D}_{H_2}\mu_{H_2}+
6(\Gamma_y+\Gamma_{yA})[\mu_{H_2}+\mu_{Q_3}+\mu_T]
+2\Gamma_{\lambda}[\mu_{\tilde s}+\mu_{H_1}+\mu_{H_2}]
\hspace{1.4cm}\nonumber\\
+2\Gamma_{hf}(\mu_{H_1}+\mu_{H_2})+2\Gamma_{H_2}\mu_{H_2}=S_{H_2}
\hspace{.4cm}\\[.1cm]
-k_{\tilde s}{\cal D}_{\tilde s}\mu_{\tilde s}+
2\Gamma_{\lambda}[\mu_{\tilde s}+\mu_{H_1}+\mu_{H_2}]+
\Gamma_{\tilde s}\mu_{\tilde s}=S_{\tilde s}\hspace{.1cm}
\end{eqnarray}
where ${\cal D}_i\equiv D_i\frac{d^2}{d\bar z^2}+v_w\frac{d}{d\bar z}$.
The transport equations
can be further simplified if the top Yukawa interactions are assumed to be
in equilibrium, which implies $\mu_{H_2}+\mu_{Q_3}+\mu_T=0$ and   
$\mu_{H_1}-\mu_{Q_3}-\mu_T=0$. We then find \cite{HS3}
\begin{eqnarray}
-k_{Q_3}{\cal D}_q\mu_{Q_3}-k_H{\cal D}_h\mu_H+
(6\Gamma_m+2\Gamma_H)\mu_H
+6\Gamma_{ss}[c_Q\mu_{Q_3}-c_H\mu_H] 
&=&S_{Q_3}+S_H
\nonumber \\[.1cm]
-(k_{Q_3}+k_T){\cal D}_q\mu_{Q_3}+k_T{\cal D}_q\mu_H
+3\Gamma_{ss}[c_Q\mu_{Q_3}-c_H\mu_H] 
&=&0
\end{eqnarray}
In this approximation the 
chargino source terms enter only via $S_H=S_{H_1}-S_{H_2}$.
Therefore, the dominating, $\theta$-dependent part of the chargino
source term (``helicity part'') cancels. The singlino, with a potentially 
large source term, decouples from the transport equations.
With the dispersion relation (\ref{dis}) the ``flavor'' part survives. Thus,
in the MSSM case  the $\delta\beta$ suppression of the baryon asymmetry is recovered.
In the kinetic momentum approach also this contribution
vanishes. Giving up the top Yukawa coupling equilibrium one also obtains a
$S_{H_1}+S_{H_2}$ contribution. In our (preliminary ) studies this still leads a
to sizable baryon asymmetry.

We solve set of diffusion equations by the Greens function method. The
weak sphalerons, which are not in equilibrium, generate the baryon 
to entropy ratio in the hot phase
\begin{equation}
\eta_B\equiv \frac{n_{\cal B}}{s}=\frac{135 \Gamma_{ws}}{2\pi^2g_*v_wT}
\int_0^{\infty}d\bar z \mu_{{\cal B}_L}(\bar z),
\end{equation}
where $\mu_{{\cal B}_L}(=7\mu_{Q_3}-2\mu_H)$ is the chemical
potential for the left-handed quark number (in the massless
approximation). 

\begin{figure}[t] 
\begin{picture}(100,85)
\put(-65,-330){\epsfxsize13cm \epsffile{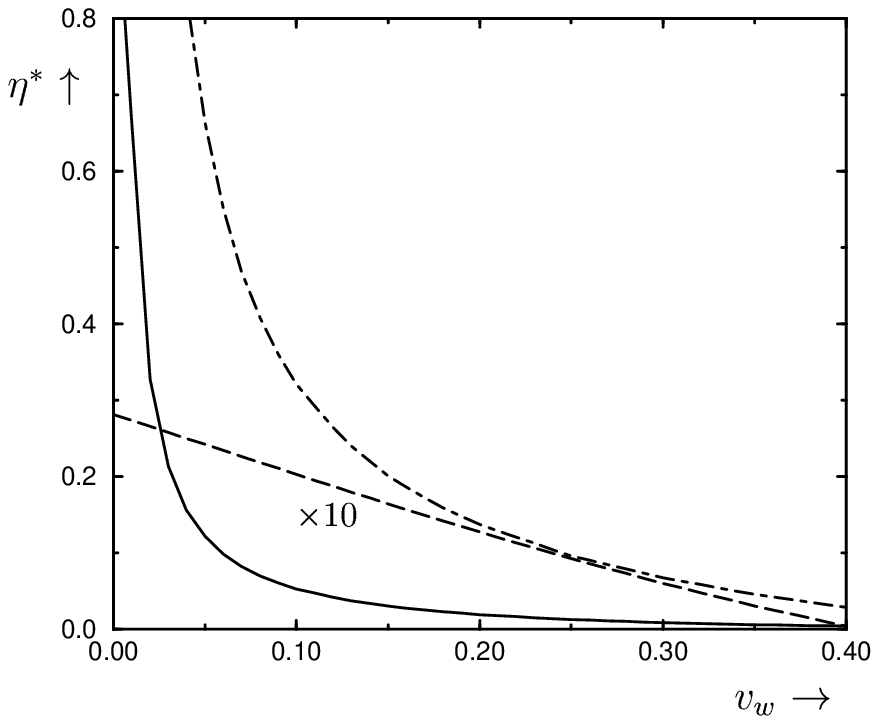}}
\put(130,-330){\epsfxsize13cm  \epsffile{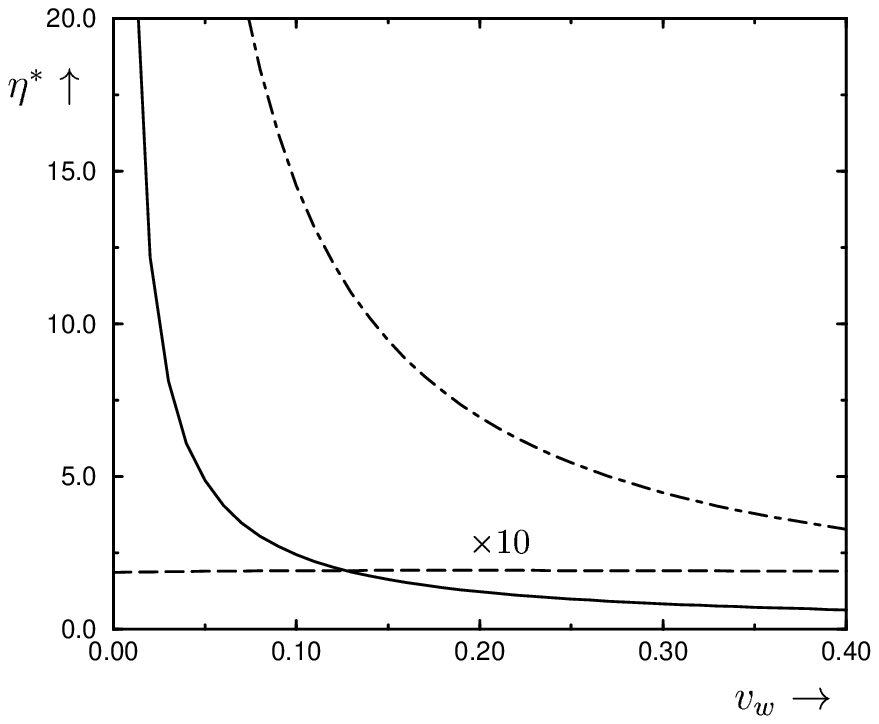}}
\put(120,120){(a)} 
\put(315,120){(b)} 
\end{picture} 
\caption{Chargino contribution to the baryon asymmetry in
units of $2\times10^{-11}$ for an example of (a): explicit CP violation (arg($\mu$)=0.1), and 
(b): transitional CP violation. The different curves correspond to
different squark spectra.
}
\label{f_3}
\end{figure}

The generated baryon asymmetry is rather sensitive to the squark spectrum.
For universal squark masses there is a large suppression by strong sphalerons. 
The baryon asymmetry increases for with $1/v_w$ (at $v_w\sim
0.01$ this behavior would be cut off by the (neglected)
effects of weak sphalerons \cite{CJK2}).
Thinner bubble walls enhance the baryon asymmetry,  $\eta \sim 1/L_w^2$.
The chargino contribution dominates the baryon production in the NMSSM.
It is especially large for $M_2\sim \mu$.
In fig.~\ref{f_3} we present two examples for the chargino contribution to
the baryon asymmetry. In the case of explicit CP-violation 
small complex phases  of the order $10^{-2}$ can account for
the observed baryon asymmetry only for very small wall velocities,
$v_w\lsim 0.01$ (and only the right-handed stop is light). However,
wall velocities in this range have recently been found in the MSSM \cite{wall}. 
In the case of transitional CP violation
a sufficient baryon number can be easily produced, also
for larger wall velocities. Hence, transitional CP violation is particularly
interesting for electroweak baryogenesis.

\noindent{\bf Acknowledgements} We would like to thank P.~John for very useful
discussions.


\begin{thebibliography}{99}

\bibitem{litestop}
J.R.~Espinosa, {\em Nucl. Phys.} {\bf B475} (1996)  273; 
B{\"o}deker, John, Laine, Schmidt, {\em Nucl.  Phys.} {\bf B497} (1997) 387; 
Carena, Quir\'os, Wagner, {\em Nucl. Phys.} {\bf  B524} (1998) 3; 
Losada, {{\tt hep-ph/9905441}}; 
Laine, Rummukainen, {\em Nucl. Phys.} {\bf B535} (1998) 423; 
ibid., {\em Phys. Rev. Lett.} {\bf 80} (1998) 5259.

\bibitem{PDFM} M.~Pietroni, {\em Nucl. Phys.} {\bf B402} (1993) 27;
                        A.T.~Davis, C.D.~Froggatt, R.G.~Moorhouse,
                         {\em Phys. Lett.} {\bf B372} (1996) 88;

\bibitem{HS2} S.J.~Huber, M.G.~Schmidt, {\em Eur. Phys. J.} {\bf C10}
                     (1999) 473.

\bibitem{HS3}S.J.~Huber, M.G.~Schmidt, hep-ph/0003122.

\bibitem{NMSSM} J.~Gunion, H.E.~Haber, G.L.~Gordon, S.~Dawson,
                           {\em The Higgs Hunters Guide}, Addison-Wesley, Reading MA, 1990.
   
\bibitem{P} P.~John, hep-ph/0010277.

\bibitem{John4}
S.J.~Huber, P.~John, M.~Laine, M.~Schmidt, Phys.\ Lett.\ {\em B475} (2000) 104.

\bibitem{CJK1} J.M.~Cline, M.~Joyce, K.~Kainulainen, {\em Phys. Lett.} {\bf B417} (1998) 79.

\bibitem{CJK2} J.M.~Cline, M.~Joyce, K.~Kainulainen, {\em JHEP} {\bf 0007} (2000) 18.

\bibitem{wall} G.~Moore, JHEP 0003:006,2000; P.~John and M.G.~Schmidt, hep-ph/0002050.

\end{thebibliography}
\end{document}